\documentclass[10pt, a4paper]{article}

\usepackage{amsmath}
\usepackage{url}
\usepackage{listings}
\usepackage{xcolor}
\usepackage{booktabs}
\usepackage[final]{lrec2026} 

\title{Quid est VERITAS?\\ A Modular Framework for Archival Document Analysis}

\name{%
\begin{tabular}[t]{c}
Leonardo Bassanini$^{1}$, Ludovico Biancardi$^{1}$, Alfio Ferrara$^{2}$,\\ 
Andrea Gamberini$^{3}$, Sergio Picascia$^{4\ast}$, Folco Vaglienti$^{3}$
\end{tabular}
\\
}

\address{$^{1}$Università degli Studi di Milano, University Library Service \\
         Via Santa Sofia, 7/9 - 20122 Milano (Italy) \\
         leonardo.bassanini@unimi.it, ludovico.biancardi@unimi.it
         \and
         $^{2}$Università degli Studi di Milano, Department of Literary Studies, Philology and Linguistics \\
         Via Festa del Perdono, 7 - 20122 Milano (Italy) \\
         alfio.ferrara@unimi.it
         \and
         $^{3}$Università degli Studi di Milano, Department of Historical Studies \\
         Via Festa del Perdono, 7 - 20122 Milano (Italy) \\
         andrea.gamberini@unimi.it, folco.vaglienti@unimi.it
         \and
         $^{4}$Università degli Studi di Milano, Department of Computer Science \\
         Via Celoria, 18 - 20133 Milano (Italy) \\
         sergio.picascia@unimi.it}

\abstract{
The digitisation of historical documents has traditionally been conceived as a process limited to character-level transcription, producing flat text that lacks the structural and semantic information necessary for substantive computational analysis. We present VERITAS (Vision-Enhanced Reading, Interpretation, and Transcription of Archival Sources), a modular, model-agnostic framework that reconceptualises digitisation as an integrated workflow encompassing transcription, layout analysis, and semantic enrichment. The pipeline is organised into four stages—Preprocessing, Extraction, Refinement, and Enrichment—and employs a schema-driven architecture that allows researchers to declaratively specify their extraction objectives. We evaluate VERITAS on the critical edition of Bernardino Corio's \textit{Storia di Milano}, a Renaissance chronicle of over 1,600 pages. Results demonstrate that the pipeline achieves a 67.6\% relative reduction in word error rate compared to a commercial OCR baseline, with a threefold reduction in end-to-end processing time when accounting for manual correction. We further illustrate the downstream utility of the pipeline's output by querying the transcribed corpus through a retrieval-augmented generation system, demonstrating its capacity to support historical inquiry.
 \\ \newline  }
 
\begin{document}

\maketitleabstract

\section{Introduction}
\label{sec:introduction}

The digitisation of historical documents has long been conceived as a process whose primary objective is the faithful transcription of textual content into machine-readable form. Under this paradigm, the output is flat text: a character sequence suitable for keyword search but lacking the structural, semantic, and relational information necessary for substantive scholarly inquiry. This limitation stems from an entrenched architectural assumption that semantic enrichment constitutes a separate, downstream task. In practice, this separation means that the majority of digitised collections never receive such enrichment, as the resources required to revisit already-processed materials and execute enrichment pipelines are rarely available. The consequence is a widening gap between the volume of digitised material and the volume that is genuinely searchable, interoperable, and amenable to computational analysis.

We argue that overcoming this gap requires reconceptualising digitisation itself: rather than treating semantic enrichment as optional post-processing, it should be an integral component of the digitisation workflow, producing structured, semantically annotated digital objects. The feasibility of this shift is enabled by recent advances in foundation models, particularly vision-language models (VLMs) and large language models (LLMs). Unlike traditional OCR systems, which operate at the character level with no awareness of document semantics, VLMs can jointly process visual and textual information, performing layout analysis, transcription, and content interpretation within a unified inference pass. LLMs, in turn, can be leveraged for downstream semantic tasks previously requiring dedicated, task-specific systems. Together, these technologies make it viable to construct pipelines traversing the full arc from raw image to semantically enriched digital object in a single automated workflow.

In this paper, we present VERITAS (Vision-Enhanced Reading, Interpretation, and Transcription of Archival Sources), a modular framework that operationalises this integrated approach. VERITAS is organised into four sequential stages, each constituting a self-contained module with well-defined inputs and outputs. The framework is model-agnostic: individual components can be substituted as newer models become available without architectural modifications. A schema-driven extraction process allows researchers to configure the target data structure according to their analytical objectives. We evaluate VERITAS on Bernardino Corio's \textit{Storia di Milano}, a Renaissance chronicle of over 1,600 pages, demonstrating substantial improvements over a commercial OCR baseline in both transcription accuracy and processing efficiency. Furthermore, we illustrate downstream utility by submitting the complete transcription to a retrieval-augmented generation (RAG) system and posing historically motivated research questions.

\section{Related Work}
\label{sec:related-work}

The automated processing of historical documents has attracted growing attention at the intersection of computer science and the digital humanities. We review the relevant literature across two areas: transcription technologies and integrated digitisation pipelines.

\subsection{Transcription of Historical Documents}

The transcription of historical documents has evolved through several technological paradigms. Commercial OCR engines such as Tesseract~\citep{smith2007tesseract} have long served as default tools; however, their reliance on character-level pattern matching renders them ill-suited to the degraded image quality, irregular layouts, and archaic typographic conventions of historical materials. Deep learning substantially improved capabilities for such documents. Transkribus~\citep{kahle2017transkribus, muehlberger2019transforming}, developed within the EU-funded READ project, became one of the most widely adopted platforms for automatic text recognition in the humanities. In parallel, eScriptorium~\citep{kiessling2019escriptorium} extended recognition to non-Latin and bidirectional writing systems. While both platforms significantly lowered the barrier of entry, they remain primarily interactive tools requiring manual ground-truth preparation, model training, and post-hoc correction.

Generative AI has opened new avenues for historical document transcription. One line of research explores LLMs as post-OCR correctors: \citet{thomas-etal-2024-leveraging} demonstrated that instruction-tuned Llama~2 models can achieve a 54.5\% reduction in character error rate on 19th-century British newspapers, though subsequent studies reported mixed results in multilingual contexts \citep{kanerva-etal-2025-ocr, boros-etal-2024-post}. A more recent line of research has shifted focus to direct transcription through vision-language models (VLMs), which jointly process visual and textual information end-to-end without a separate OCR stage. Several evaluations demonstrate that VLMs can match or surpass dedicated OCR systems: \citet{humphries2024unlockingarchivesusinglarge} reported character error rates of 5--7\% on 18th--19th century English manuscripts, \citet{kim2025earlyevidencellmsoutperform} found general-purpose VLMs outperforming traditional OCR on historical tabular documents, and \citet{levchenko2025evaluatingllmshistoricaldocument} benchmarked 12 multimodal LLMs on 18th-century Russian texts, highlighting both promise and pitfalls such as \textit{over-historicisation}. On the specialised side, CHURRO~\citep{semnani2025churro}, a 3B-parameter VLM fine-tuned on 155 historical corpora spanning 46 language clusters, demonstrated that targeted training yields accuracy superior to commercial alternatives at a fraction of the cost. VERITAS builds on this latter paradigm, employing VLMs as its primary extraction mechanism while retaining flexibility for LLM-based refinement.

\subsection{Integrated Digitisation Pipelines}

While significant progress has been made on individual components, fewer efforts have addressed their integration into end-to-end pipelines. The OCR-D project~\citep{neudecker2019ocrd} developed a modular framework for OCR processing of historical printed documents in German libraries, emphasising interoperability through standardised formats, though its scope does not extend to semantic enrichment. The DAHN project~\citep{Chiffoleau2024} proposed a TEI-centred pipeline comprising six stages, leveraging eScriptorium for HTR and TEI Publisher for dissemination, representing an important step but remaining oriented towards digital scholarly editions rather than flexible data extraction. At larger scales, SocFace~\citep{boillet2024socface} demonstrated end-to-end pipelines for French censuses, and \citet{constum2024endtoendinformationextractionhandwritten} proposed an approach for handwritten Parisian marriage records. These efforts, however, tend to be tightly coupled to specific document genres.

VERITAS extends this body of work in several respects. Unlike interactive platforms such as Transkribus and eScriptorium, it operates as a fully automated, configurable pipeline. Unlike OCR-focused frameworks such as OCR-D, it extends beyond transcription to encompass semantic enrichment, entity linking, and structured data indexing. Its model-agnostic architecture provides flexibility absent from existing designs. Finally, by demonstrating downstream utility for LLM-assisted scholarly inquiry (Section~\ref{sec:downstream}), we address a gap where evaluation typically stops at transcription accuracy without assessing usability for substantive research.

\section{Methodology}
\label{sec:methodology}

\begin{figure*}[ht!]
    \centering
    \includegraphics[width=0.9\linewidth]{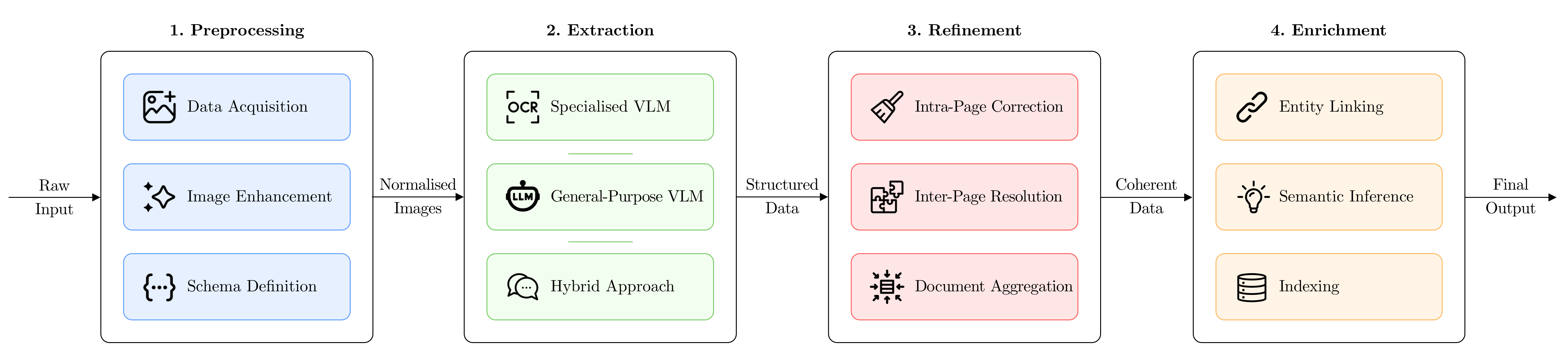}
    \caption{VERITAS architecture diagram illustrating the four stages and their inputs and outputs.}
    \label{fig:pipeline}
\end{figure*}

VERITAS (Vision-Enhanced Reading, Interpretation, and Transcription of Archival Sources) is designed to transform raw historical documents into structured, semantically enriched, machine-readable data. By leveraging VLMs and LLMs, the pipeline extends digitisation beyond character-level transcription towards comprehensive document understanding. The architecture is model-agnostic, allowing practitioners to substitute models optimised for particular languages or document types without architectural modifications. Combined with schema-driven extraction, this ensures broad applicability across diverse archival collections.

The pipeline comprises four sequential stages—Preprocessing, Extraction, Refinement, and Enrichment—each a self-contained module with well-defined inputs and standardised outputs (Figure~\ref{fig:pipeline}). Individual components can be activated, bypassed, or substituted according to project requirements. Throughout this section, we illustrate operations using pages from the case study in Section~\ref{sec:evaluation}: the critical edition of Bernardino Corio's \textit{Storia di Milano} (1978), a Renaissance chronicle of over 1,600 pages.

\subsection{Preprocessing}

The Preprocessing stage transforms heterogeneous raw inputs into a standardised format suitable for automated analysis through three operations.

\textbf{Data Conversion}. The pipeline accepts input formats commonly encountered in archival research, e.g. digitised PDFs, individual page scans (TIFF, JPEG, PNG), and photographs captured with handheld devices, and converts them into a uniform representation, ensuring that all subsequent pipeline stages operate on a homogeneous, high-quality image format. Each input is rendered as a high-resolution raster image normalised to a consistent colour space and resolution.

\begin{figure}[ht!]
    \centering
    \includegraphics[width=\linewidth]{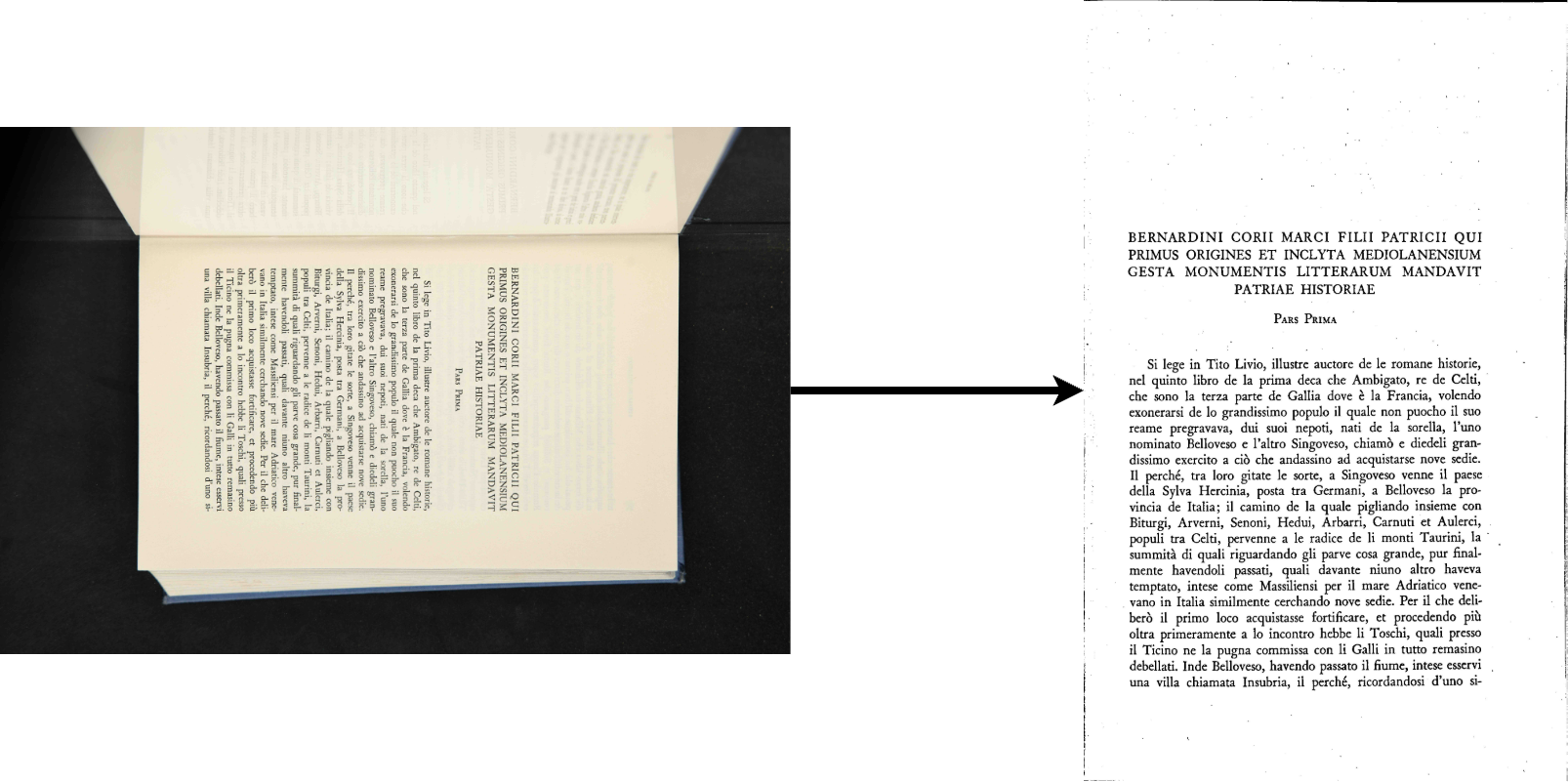}
    \caption{Preprocessing operations applied to a page from the critical edition of Corio's \textit{Storia di Milano}. Left: the original colour scan, captured in landscape orientation with portions of the adjacent page visible. Right: the result after rotation correction, page detection, grayscale conversion, and adaptive thresholding.}
    \label{fig:preprocessing}
\end{figure}

\textbf{Image Enhancement}. Optional computer vision techniques may be applied to the normalised images: deskewing (correcting rotational misalignment), denoising (removing noise, stains, or scanning artifacts), binarisation (converting to black-and-white to increase text-background contrast), and page detection (isolating the document area from extraneous background elements). The specific enhancement operations applied are configurable based on the quality and characteristics of the source materials. Figure~\ref{fig:preprocessing} shows the result of these operations conducted on an image from the proposed case study.

\textbf{Schema Definition}. This operation constitutes the critical interface between scholarly intent and machine-processable output: analytical requirements are formalised into a structured specification determining what information the pipeline extracts and in what form. The schema, typically expressed in JSON Schema, defines the semantic fields to populate for each document element, and determines the tools employed in subsequent stages. Its design is a collaborative exercise between domain experts, who articulate driving research questions, and technical personnel, who translate these into formal specifications. Below is an illustrative example.

\begin{lstlisting}
{
    "type": "object",
    "properties": {
        "bbox": {
            "type": "array",
            "items": { "type": "integer" }
    },
        "category": {
            "type": "string",
            "enum": ["title", "text", "header",
                    "footnote", "figure", "table"]
        },
        "text": { "type": "string" },
        "speaker": { "type": "string" },
        "date": { "type": "string" },
        "place": { "type": "string" },
        "entities": {
            "type": "array",
            "items": {
                "type": "object",
                "properties": {
                    "mention": { "type": "string" },
                    "type": { 
                        "type": "string",
                        "enum": ["person", "institution", "place"]
                    }}}}},
    "required": ["bbox", "category", "text"]
}
\end{lstlisting}

\subsection{Extraction}

The Extraction stage constitutes the core of the pipeline, performing the transformation of visual information into machine-readable structured data. This stage is designed with high modularity, offering three distinct processing paths that can be selected based on project requirements, available computational infrastructure, and the complexity of the target extraction tasks.

\begin{itemize}
    \item \textbf{Specialised Vision-Language Models}. Compact, resource-efficient VLMs optimised for document analysis, achieving high accuracy in layout detection, text localisation, reading order determination, and transcription. Their advantage lies in computational efficiency and transcription fidelity; however, they operate according to a fixed output schema and cannot accommodate arbitrary user-defined instructions.
    \item \textbf{General-Purpose Vision-Language Models}. Large-scale VLMs capable of processing both visual and textual information. While potentially exhibiting marginally lower performance on specialised document analysis benchmarks compared to the previous models, their strength resides in flexibility: users provide natural language instructions to guide extraction towards specific information needs beyond conventional transcription.
    \item \textbf{Multi-Step Hybrid Approach}. Combines both approaches through a two-phase process: a specialised VLM performs foundational layout detection and transcription, then a general-purpose LLM or VLM refines and enhances extraction according to user-specified instructions. This path suits complex documents requiring both accurate transcription and sophisticated semantic interpretation.
\end{itemize}

\begin{figure}[ht!]
    \centering
    \includegraphics[width=\linewidth]{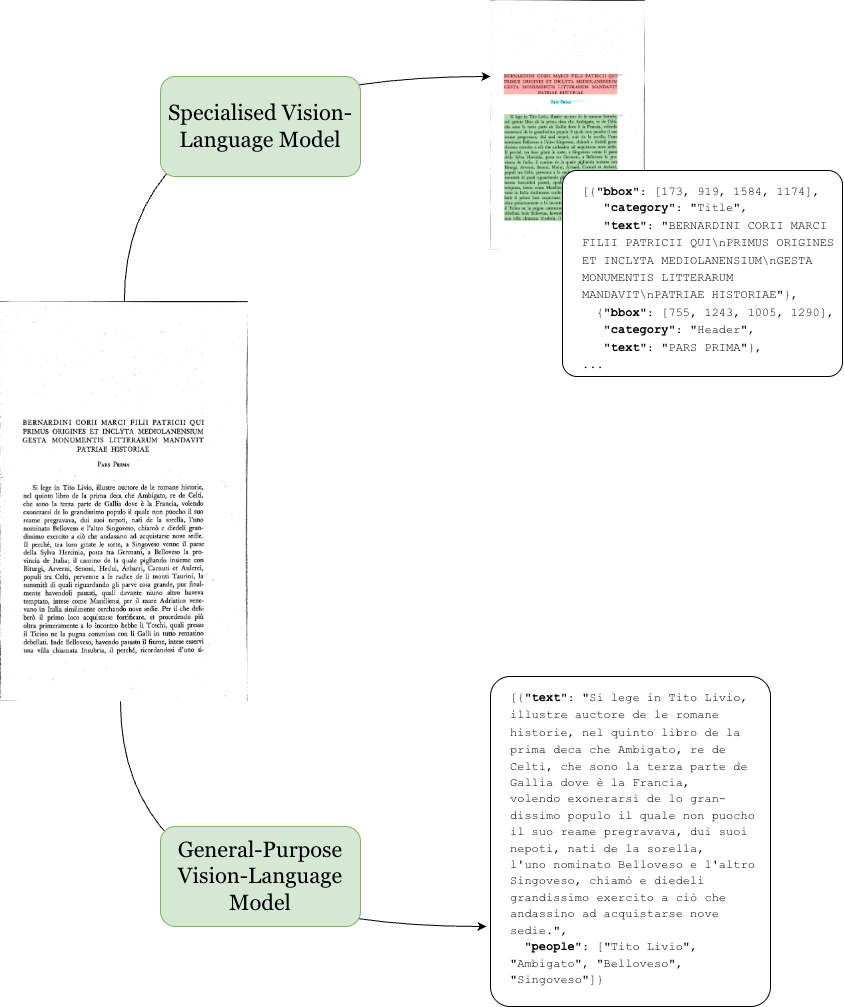}
    \caption{The difference in output between the Specialised Vision-Language Model and the General-Purpose Vision-Language Model processing paths.}
    \label{fig:extraction}
\end{figure}

The selection among these paths is determined by the interplay of several factors, including the available computational resources, the quality and complexity of the source documents, and the depth of semantic analysis required. For projects prioritising transcription accuracy with minimal computational overhead, the first path offers an efficient solution. For projects requiring flexible, instruction-guided extraction or semantic inference, the second and third path are more appropriate. Figure~\ref{fig:extraction} contrasts the predefined output of a specialised VLM against the more flexible output of a general-purpose VLM performing semantic inference.

\subsection{Refinement}

The Refinement stage transforms the collection of page-level extraction outputs into a single, coherent document representation through a series of cleaning, consolidation, and aggregation operations.

\textbf{Intra-Page Correction}. This operation addresses artifacts and inconsistencies within individual page outputs. Typical corrections include the reconstruction of hyphenated words split across line breaks, normalisation of typographic conventions (e.g., unifying quotation mark styles, standardising whitespace), and validation of the extracted data against the defined schema to ensure structural integrity.

\textbf{Inter-Page Resolution}. This operation resolves dependencies and continuities that span page boundaries. In historical documents, content units frequently extend across multiple pages, e.g a newspaper article beginning on one page and concluding on another, or a parliamentary speech spanning several folios. Inter-page resolution ensures that such fragmented content is correctly identified and linked, and that metadata is propagated appropriately across page breaks.

\textbf{Document Aggregation}. The refined page-level outputs are merged into a unified data structure representing the complete source document. This aggregation produces a single coherent object that serves as input to the subsequent Enrichment stage.

\subsection{Enrichment}

The Enrichment stage enhances the value of the structured data by connecting it to external knowledge sources, obtaining additional semantic insights, and preparing it for storage and downstream analysis.

\textbf{Entity Linking}. Named entities identified within the transcribed text—such as persons, organisations, and locations—are disambiguated and linked to canonical identifiers in external knowledge bases (e.g., Wikidata, VIAF, or domain-specific authority files). This process not only resolves ambiguities (e.g., distinguishing between individuals sharing the same name) but also situates the document within a broader knowledge graph, enabling cross-referencing and relational queries.

\textbf{Semantic Inference}. This operation leverages LLMs to perform advanced analytical tasks and infer information not explicitly present in the source text. Depending on the research objectives, such tasks may include topic classification, sentiment analysis, event extraction, temporal reasoning, named entity recognition for domain-specific entity types, or abstractive summarisation. The schema-driven architecture of the pipeline allows these inferred annotations to be systematically incorporated into the structured output. These operations can be also performed with general-purpose VLMs during the Extraction stage.

\textbf{Indexing}. The indexing operation formats the enriched data according to established standards appropriate to the target research community and loads it into suitable storage systems. For humanities scholarship, standard serialisation formats such as XML-TEI (Text Encoding Initiative) facilitate interoperability with existing digital humanities infrastructures. For quantitative analysis, tabular formats (e.g., CSV) or integration with database systems (e.g., relational databases, full-text search engines, vector or graph databases) enable efficient querying and statistical processing.

\begin{figure}[ht!]
    \centering
    \includegraphics[width=0.9\linewidth]{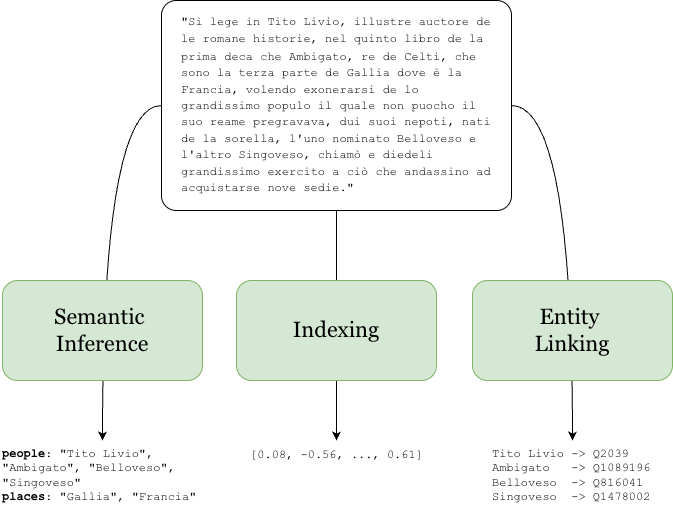}
    \caption{Examples of Enrichment operations.}
    \label{fig:enrichment}
\end{figure}

Figure~\ref{fig:enrichment} shows different examples of Enrichment operations: through semantic inference we can extract named entities, e.g. people and places; an encoder can compute the embedding of a textual element to index it for semantic search; we can match the entities within the text with external knowledge bases.

\section{Evaluation}
\label{sec:evaluation}

To assess the effectiveness of the VERITAS pipeline, we conducted an empirical evaluation on a representative historical document collection. This evaluation focused on two primary dimensions: (i) transcription accuracy, measured through standard OCR quality metrics, and (ii) computational efficiency, assessed via processing time analysis. Additionally, we evaluated the pipeline's capacity for accurate layout analysis and element extraction.

\subsection{Case Study}

The evaluation was conducted on the \textit{Storia di Milano} (\textit{History of Milan}), a Renaissance chronicle authored by Bernardino Corio (1459--c.~1519). The digitisation of this work was undertaken as part of a broader interdisciplinary project at our university, bringing together historians, computational linguists, and computer scientists with the shared objective of making this primary source accessible for large-scale computational analysis. Corio's work constitutes one of the most significant sources for the history of the Duchy of Milan, spanning from antiquity to the late fifteenth century. For this study, we employed the critical edition curated by Anna Morisi Guerra, published by UTET (Turin) in 1978 as part of the \textit{Classici della Storiografia} series~\citep{corio1978storia}. The edition comprises two volumes totalling 1,688 pages.

Although this critical edition employs modern typography, thereby avoiding the challenges posed by historical typefaces or manuscript hands, it nonetheless presents a rich variety of difficulties for automated processing that motivated its selection as an evaluation testbed. The text alternates between vernacular and modern Italian, requiring the extraction system to handle two distinct languages within a single document flow. The page layouts are heterogeneous, comprising dense prose, extensive footnotes, illustrative figures, tables, indices, and front matter, each demanding correct identification and classification by the layout analysis component.

\subsection{Evaluation Methodology}

\textbf{Ground Truth Construction}. A subset of 100 pages was selected from the corpus using stratified sampling to ensure adequate representation of the diverse page layouts present in the edition (e.g., pages with dense prose, pages with extensive footnotes, title pages, and pages containing illustrations or tables). Three domain experts, working from a shared set of transcription guidelines, manually transcribed these pages to establish a reference ground truth. Additionally, the structural elements identified by the vision-language model (text, headers, footnotes, etc.) were manually verified and corrected by the same annotators to provide ground truth for layout analysis evaluation.

\textbf{Baseline}. To contextualise the performance of the VERITAS pipeline, we compared it against ABBYY FineReader, a widely adopted commercial OCR solution that is also the standard digitisation tool currently employed by our university library. The software was executed with default configuration settings, which is appropriate given that the critical edition employs modern typefaces that do not require specialised historical document profiles.

\textbf{Pipeline Configuration}. For this evaluation, the VERITAS pipeline was configured as follows. In the preprocessing stage, several image normalisation operations were applied to address artifacts introduced during digitisation. The original scans were captured in landscape orientation with portions of adjacent pages partially visible in the frame; furthermore, the scans were acquired in colour. To prepare the images for extraction, we applied rotation correction to restore portrait orientation, employed an object detection model to identify and isolate the region of interest corresponding to the target page~\citep{doc_ufcn2021}, and converted the images to grayscale. Additionally, adaptive thresholding techniques were applied to enhance text-background contrast and improve content legibility. For the extraction stage, the pipeline was configured to use Path A (Specialised Vision-Language Models), employing a document-specialised VLM, specifically, \texttt{dots.ocr}~\citep{li2025dotsocrmultilingualdocumentlayout}, followed by minimal post-processing operations consisting of end-of-line hyphenation correction and output format normalisation (removal of markdown artefacts).

\textbf{Metrics}. Transcription quality was assessed using two standard metrics: Word Error Rate (WER), defined as the minimum number of word-level insertions, deletions, and substitutions required to transform the predicted transcription into the ground truth, normalised by the total number of words in the reference; and Character Error Rate (CER), analogously defined at the character level, providing a finer-grained assessment of transcription fidelity. Both metrics were computed at the corpus level, i.e., over the concatenated text of all evaluated pages, to avoid potential bias introduced by page-length variability. To isolate substantive transcription errors from superficial formatting differences, we report results under two conditions: Raw (no normalisation) and Normalised (text converted to lowercase with punctuation removed). We also evaluated element extraction computing the F1 Score: an extracted element was considered a true positive if it matched a ground-truth element in both spatial location (bounding box overlap) and semantic label; false positives comprised spurious detections or incorrect label assignments; false negatives represented missed elements.

\subsection{Results}

Table~\ref{tab:ocr_results} presents the corpus-level transcription error rates for both systems. The results demonstrate that the VLM-based approach employed in VERITAS consistently outperforms the commercial baseline across all metrics and normalisation conditions. Under normalised evaluation, VERITAS achieves a WER of 1.1\% and a CER of 0.7\%, representing relative improvements of 67.6\% and 50.0\%, respectively, compared to ABBYY FineReader. The performance gap is even more pronounced under raw evaluation conditions.

\begin{table}[ht!]
    \centering
    \begin{tabular}{lcccc}
    \toprule
        & \multicolumn{2}{c}{\textbf{ABBYY}} & \multicolumn{2}{c}{\textbf{VERITAS}} \\
    \cmidrule(lr){2-3} \cmidrule(lr){4-5}
    \textbf{Text} & WER & CER & WER & CER \\
    \midrule
    Raw        & 0.142 & 0.031 & \textbf{0.055} & \textbf{0.013} \\
    Normalised & 0.034 & 0.014 & \textbf{0.011} & \textbf{0.007} \\
    \bottomrule
    \end{tabular}
    \caption{Corpus-level transcription error rates computed over the concatenated text of all 100 evaluated pages. Lower values indicate better performance.}
    \label{tab:ocr_results}
\end{table}

Regarding processing efficiency, while the VLM-based approach incurs substantially higher computational cost per page when processed individually (46 seconds versus 4 seconds for ABBYY), modern inference frameworks enable efficient concurrent processing through automatic request batching. In our experimental configuration, we employed vLLM~\citep{kwon2023efficient}, a high-throughput serving framework that dynamically batches incoming requests and optimises GPU memory utilisation. Using one third of the memory allocation of a single Nvidia H100 NVL 94Gb GPU, vLLM's continuous batching mechanism reduced the effective per-page processing time to 0.89 seconds, a 4.5$\times$ improvement over the commercial solution. Beyond automated processing time, however, the practical impact of the pipeline must also account for the manual effort required to correct its output. The domain experts recorded the time needed to correct the ABBYY FineReader transcriptions, yielding a mean correction time of 2 minutes and 15 seconds per page. Extrapolated to the full 1,688-page corpus, this amounts to approximately 63 hours of manual post-correction labour. Since VERITAS reduces the normalised WER by 67.6\% relative to ABBYY, a proportional reduction in correction effort can be reasonably assumed, bringing the estimated per-page correction time to approximately 44 seconds and the projected corpus-level total to roughly 20 hours, a saving of over 40 hours of expert labour. Combining automated processing and manual correction, the total estimated time for producing a verified transcription of the full edition decreases from approximately 65 hours with ABBYY to approximately 21 hours with VERITAS, a threefold reduction in end-to-end effort. These gains should, however, be interpreted in light of the computational profile of the selected extraction model. While concurrent inference substantially lowers effective per-page latency, VLM-based processing remains more demanding than conventional OCR in terms of GPU memory, serving infrastructure, and energy consumption.

Finally, the element extraction performance of the VLM-based approach confirms its reliability for layout analysis, with the model achieving an F1 Score of $0.966$. These results indicate that the specialised VLM reliably identifies and correctly classifies the structural elements present in the document pages, a capability essential for the subsequent refinement and enrichment stages of the pipeline, as accurate element extraction enables proper content aggregation and semantic annotation.

\section{Downstream Application}
\label{sec:downstream}

The preceding evaluation demonstrates that the VERITAS pipeline achieves high-fidelity transcription and reliable layout analysis. However, the ultimate value of a digitisation framework for the humanities resides not merely in the accuracy of its output, but in the degree to which that output can support substantive scholarly inquiry. To illustrate this potential, we conducted an exploratory study in which the complete transcribed text of the \textit{Storia di Milano}, produced by the VERITAS pipeline, was ingested into a retrieval-augmented generation (RAG) system, and a set of historically motivated research questions was posed to the model.

\subsection{Setup}

On top of the indexing results of the Enrichment phase of VERITAS, we developed a dedicated RAG pipeline tailored to the structure of Corio's chronicle. The 1,688 digitised pages were processed by a custom ingestion module that annotates each page with temporal and structural metadata and groups them into chronologically coherent chunks aligned with the chronicle's year markers and chapters divisions. The resulting corpus comprises approximately 1,331 content chunks and 200 footnotes, embedded using BAAI/bge-m3\footnote{\url{https://huggingface.co/BAAI/bge-m3}}~\citep{bge-m3} and indexed in a ChromaDB collection with cosine distance.

At query time, an embedding-based router classifies each question as either \textit{specific} (targeting events, persons, or dates) or \textit{general} (spanning themes, style, or interpretive questions). Specific queries trigger year-filtered semantic search with footnote augmentation, while general queries employ Maximal Marginal Relevance (MMR) reranking to ensure temporal and thematic diversity across retrieved chunks. Retrieved passages are then fed to a generative model, GLM-4.7-Flash\footnote{\url{https://huggingface.co/zai-org/GLM-4.7-Flash}}~\cite{5team2025glm45agenticreasoningcoding}, with prompt templates that enforce grounding in the source material and instruct the model to match the query language.

A panel of historians formulated a set of research questions spanning diverse analytical dimensions: factual retrieval, interpretive analysis, thematic synthesis, prosopographic reconstruction, and cross-referencing of events and actors. These questions were designed to reflect the types of inquiry that scholars would naturally pursue when working with this primary source. From this set, we selected three representative questions for detailed discussion, each exemplifying a distinct mode of historical inquiry.

\subsection{Representative Queries}

\textbf{Factual Entity Extraction.} The question \textit{``Who were the ducal secretaries and chancellors during the Sforza era?''} requires the model to identify specific individuals and their institutional roles from mentions dispersed across the chronicle. The system produced a well-organised response grounded in the retrieved passages. It correctly identified Cicco Simonetta as \textit{general secretary}, citing the chronicle's account of his formal appointment following the death of Galeazzo Maria Sforza in 1477, and described his sweeping administrative authority over both domestic and foreign affairs. The response further retrieved Giovanni Francesco Marliano, appointed as jurist and governor during Ludovico Sforza's departure from Italy in 1499, as well as peripheral figures such as Bernardino Curtio and his brother Iacopo, named prefect and captain of the Milanese fortresses during the Duke's illness in 1489. Notably, the system also reconstructed the institutional reorganisation into two senates described in the chronicle: one for civil affairs in the Corte dell'Arenga and another for state deliberations in the castle, where Simonetta and his associates exercised decisive influence. This response demonstrates the system's capacity to aggregate factual information scattered across hundreds of pages into a structured outcome, a task that would require considerable manual effort if conducted through traditional close reading alone.

\textbf{Interpretive Stance Detection.} The question \textit{``Does the author reveal his political sympathies?''} demands a qualitatively different analytical operation: the model must synthesise evidence of authorial bias across the entire work and articulate an interpretive judgment. The system's response identified a multi-layered political stance that evolves over the course of the chronicle. It recognised Corio's explicit loyalty to the Sforza dynasty. The response detected an early apologetic posture, exemplified by the author's praise of Francesco Sforza's restoration of Milan's fortifications and his justification of ducal authority as a bulwark against popular disorder. However, the system also identified a progressive disenchantment in the later portions of the work: Corio's depiction of Giovanni Galeazzo Maria as a ruler corrupted by ministerial avarice, and his characterisation of \textit{hybris} as the root cause of the Sforza downfall, reveal a capacity for self-critical judgment regarding his own patrons. This response illustrates the system's ability to move beyond literal extraction toward historical interpretation, identifying ideological tensions within the source.

\textbf{Thematic Synthesis and Causal Reasoning.} The question \textit{``Which epidemics does the author record and what sanitary measures were taken to limit contagion?''} requires the model to identify a recurring thematic thread across several centuries of narrative and, for each instance, link the event to any associated policy response. The system identified multiple epidemic events, most prominently the \textit{peste acerrima} of 1485, which Corio describes as having driven him into rural retreat and which directly motivated the composition of the chronicle, and the pestilence of 1450, reported to have caused approximately thirty thousand deaths and severely disrupted the Jubilee. For each event, the model provided contextual details drawn from the source text, including Corio's autobiographical account of fleeing to the countryside. Regarding sanitary measures, the response correctly noted that the chronicle reflects predominantly reactive and informal responses: rural withdrawal as a form of quarantine, obligations placed on rulers such as Emperor Henry VII to maintain urban infrastructure at their own expense (bridges, roads), and the role of civic assemblies (\textit{Credentia}) in coordinating crisis management. This response demonstrates the system's capacity for multi-hop reasoning: extracting thematically related passages distributed across the chronicle and synthesising them into a coherent analytical account.

\subsection{Discussion}

The exploratory results presented above suggest that RAG-based LLM systems, when provided with high-quality transcriptions produced by the VERITAS pipeline, can serve as effective tools for preliminary historical analysis. However, while these results are encouraging, they should be interpreted with appropriate caution. The responses have not been subjected to systematic validation, and RAG-based systems remain susceptible to hallucination, particularly when queries require inference beyond what is explicitly stated in the source material. A rigorous evaluation of factual accuracy and interpretive validity, conducted in collaboration with domain historians, constitutes an essential direction for future work. Notwithstanding these limitations, this demonstration highlights the potential of integrating high-fidelity transcription pipelines with LLM-based querying systems to lower the barrier of entry for large-scale historical document analysis, enabling scholars to formulate and explore research hypotheses across extensive corpora more efficiently than traditional manual methods would allow.

\section{Conclusion}
\label{sec:conclusion}

We have presented VERITAS, a modular, model-agnostic framework that reconceptualises historical document digitisation as an integrated process encompassing transcription, structural analysis, and semantic enrichment within a unified pipeline. The framework's schema-driven architecture allows researchers to declaratively specify their extraction objectives, ensuring that the pipeline's outputs are tailored to the analytical needs of diverse scholarly communities.

Our evaluation on the critical edition of Corio's \textit{Storia di Milano} demonstrates that a VLM-based extraction pipeline can substantially outperform a commercial OCR baseline, achieving a 67.6\% relative reduction in word error rate under normalised conditions, while concurrent inference reduces effective per-page processing time by a factor of 4.5. When accounting for the manual correction effort that remains indispensable in any digitisation workflow, these improvements translate into an estimated threefold reduction in end-to-end processing time for the complete 1,688-page corpus. The downstream application of the pipeline's output through a RAG-based system further illustrates that high-fidelity, structured transcriptions can directly support substantive historical inquiry, from factual entity extraction to interpretive stance detection.

Several directions for future work remain. First, a systematic evaluation of the Enrichment stage, including entity linking accuracy and the reliability of LLM-based semantic inference on historical texts, is needed to validate the full pipeline beyond its transcription capabilities. Second, the downstream RAG-based analysis presented here is exploratory; a rigorous assessment of factual accuracy and interpretive validity, conducted in collaboration with domain historians, is essential. Third, we intend to evaluate VERITAS on document collections that pose greater palaeographic challenges, such as manuscript sources and early printed books with non-standard typefaces, to assess the generalisability of the approach. Finally, we plan to release the framework as an open-source toolkit to facilitate adoption and community-driven extension across the digital humanities.


\section*{Bibliographical References}
\label{sec:reference}

\bibliographystyle{lrec2026-natbib}
\bibliography{refs}



\end{document}